\documentclass[conference]{IEEEtran}
\IEEEoverridecommandlockouts
\usepackage{cite}
\usepackage{amsmath,amssymb,amsfonts}
\usepackage{algorithmic}
\usepackage{graphicx}
\usepackage{textcomp}
\usepackage{xcolor}
\usepackage{color,array}
\usepackage{orcidlink}
\usepackage{tcolorbox}
\usepackage{booktabs}
\usepackage{float}
\usepackage{amsmath} 

\def\BibTeX{{\rm B\kern-.05em{\sc i\kern-.025em b}\kern-.08em
    T\kern-.1667em\lower.7ex\hbox{E}\kern-.125emX}}

\begin{document}

\title{Next-Generation Phishing: How LLM Agents Empower Cyber Attackers\\

\thanks{\textsuperscript{$\mathsection$} J. Chen acknowledged the support by the Fordham Office of Research through a Fordham AI Research (FAIR) Grant. }
}

\author{
    Khalifa Afane\textsuperscript{*}, Wenqi Wei\textsuperscript{*}, Ying Mao\textsuperscript{*}, Junaid Farooq\textsuperscript{†}, and Juntao Chen\textsuperscript{*$ \mathsection$} \\
    \textsuperscript{*}Department of Computer and Information Sciences, Fordham University, New York, NY, USA \\
    \textsuperscript{†}Department of Electrical and Computer Engineering, University of Michigan-Dearborn, Dearborn, MI, USA \\
    \texttt{\{mafane, wenqiwei, ymao41, jchen504\}@fordham.edu}, \texttt{mjfarooq@umich.edu}
}

\maketitle

\begin{abstract}
The escalating threat of phishing emails has become increasingly sophisticated with the rise of Large Language Models (LLMs). As attackers exploit LLMs to craft more convincing and evasive phishing emails, it is crucial to assess the resilience of current phishing defenses. In this study we conduct a comprehensive evaluation of traditional phishing detectors, such as Gmail Spam Filter, Apache SpamAssassin, and Proofpoint, as well as machine learning models like SVM, Logistic Regression, and Naive Bayes, in identifying both traditional and LLM-rephrased phishing emails. We also explore the emerging role of LLMs as phishing detection tools, a method already adopted by companies like NTT Security Holdings and JPMorgan Chase. Our results reveal notable declines in detection accuracy for rephrased emails across all detectors, highlighting critical weaknesses in current phishing defenses. As the threat landscape evolves, our findings underscore the need for stronger security controls and regulatory oversight on LLM-generated content to prevent its misuse in creating advanced phishing attacks. This study contributes to the development of more effective Cyber Threat Intelligence (CTI) by leveraging LLMs to generate diverse phishing variants that can be used for data augmentation, harnessing the power of LLMs to enhance phishing detection, and paving the way for more robust and adaptable threat detection systems.
\end{abstract}

\begin{IEEEkeywords}
Large language models, Cybersecurity, Email phishing detection, Semantic evasion.
\end{IEEEkeywords}

\section{Introduction}
Phishing emails remain a prevalent and persistent threat in cybersecurity, often exploiting human psychology to deceive recipients into revealing sensitive information~\cite{almomani2013survey}, \cite{jones2019email}, \cite{mcalaney2020understanding}. Which directly led to large organizations averaging a loss of \$15 million in 2023 \cite{wang2023mitigating}.  Traditional phishing detectors have achieved high precision and recall by recognizing specific linguistic cues, effectively mitigating many phishing attempts. However, the rapid advancement of Large Language Models (LLMs) has introduced new complexities into this landscape. These sophisticated models has already outperformed domain experts on cybersecurity benchmarks like CyberMetric\cite{tihanyi2024cybermetric}, making them useful for crafting more nuanced and convincing phishing emails, rendering classical phishing datasets and detection methods increasingly less effective. As a result, LLM-generated phishing emails pose a significant new threat that needs to be urgently addressed. This challenge has been further exacerbated by advancements in prompt engineering techniques, such as zero-shot and few-shot prompting, which are effective for new tasks without requiring training data, enabling attackers to generate highly targeted and contextually accurate emails with minimum effort ~\cite{sahoo2024systematic}, \cite{radford2019language}.

Prior to the rise of LLMs significant research had already enhanced phishing detection methods. For instance, Fette et al.~\cite{fette2007learning} introduced a machine learning approach focusing on features designed to detect deception, successfully identifying over 99.5\% of phishing emails with a very low false positive rate. Ma et al.~\cite{ma2009detecting} expanded on this by using hybrid features, combining content-based keywords and phrases with attributes like forms, and mismatched URLs, to build robust classifiers. Similarly, Verma et al.~\cite{verma2012detecting} explored natural language processing techniques, highlighting the effectiveness of semantic analysis in identifying phishing attempts. Other techniques have also proven to be very effective, achieving a near-perfect accuracy ~\cite{sheng2009empirical}, ~\cite{alhogail2021applying}. 

\begin{figure*}[h!]
\centering
\includegraphics[width=1\textwidth, height=0.19\textheight]{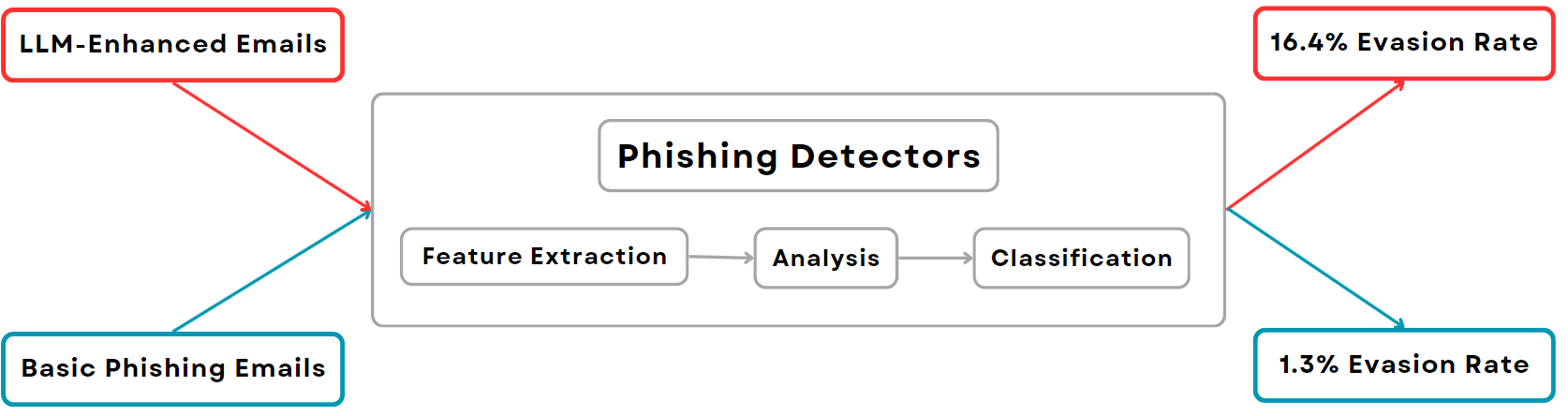}
\caption{Evaluation methodology workflow, highlighting the differences in detection effectiveness on average between traditional and LLM-rephrased emails.}
\label{fig:framework}
\end{figure*}

The dual-use nature of LLMs makes them particularly relevant in this context, as they are highly effective in generating both beneficial and malicious content. Wu et al.~\cite{wu2024new} and Yao et al.~\cite{yao2024survey} highlight security concerns posed by LLMs, noting their ability to craft sophisticated phishing emails that evade traditional detection methods. Additionally, LLMs have been shown to generate personalized phishing messages at scale, realistically and cost-effectively~\cite{hazell2023spear}. LLMs such as Llama~\cite{touvron2023llama}, Gemini~\cite{team2023gemini}, Claude~\cite{anthropic_claude}, and GPT~\cite{achiam2023gpt} models demonstrate remarkable proficiency in generating human-like text, increasingly applied to tasks like phishing detection~\cite{heiding2024devising},~\cite{koide2024chatspamdetector}. Building on prior research, this study underscores the dual nature of these models: they can craft sophisticated phishing emails while also enhancing Cyber Threat Intelligence (CTI) by improving phishing detectors through augmented data training.

The rest of the paper is organized as follows: Section II reviews related works on phishing detection and the challenges posed by LLM-generated phishing emails. Section III presents the methodology, including the datasets used, experimental setup, and rephrasing techniques. Section IV discusses the experimental results, highlighting the performance of phishing detectors and machine learning models. Section V provides a discussion of the findings, outlines limitations, and suggests future research directions. Finally, Section VI concludes the paper with a summary of key insights and contributions.

\section{Related Works}

Phishing email detection has traditionally relied on non-LLM-based methods, such as Google's built-in Gmail spam filter, other prominent tools like SpamAssassin~\cite{spamassassin} and Proofpoint have also proven to be highly effective detectors~\cite{brickley2021comparative}. However, the integration of LLM-based phishing detection is an emerging phenomenon \cite{chataut2024can}, with companies like JPMorgan Chase leading the charge~\cite{labonne2023spam}, and NTT Holdings introducing their framework ChatSpamDetector~\cite{koide2024chatspamdetector}.

However, as phishing detection improves, LLMs are being harnessed to escalate the sophistication of phishing attacks, posing new challenges. Hazell ~\cite{hazell2023spear}. for example, explored the potential of LLMs to scale spear-phishing campaigns by generating personalized emails for over 600 British Members of Parliament using GPT-3.5 and GPT-4 models. The findings show that these models can create realistic and cost-effective spear-phishing emails, with each email costing only a fraction of a cent. Similarly, Heiding et al.~\cite{heiding2024devising} investigated the use of LLMs combined with V-Triad in email phishing generation and found that models like GPT-4 when pared with human knowledge are capable of generating highly convincing phishing emails that can evade traditional detection methods. Kang et al. ~\cite{kang2023exploiting} further explored the use of LLMs in various malicious tasks, concluding that these models are capable of generating well-designed content that can be exploited for phishing and other scams.

In contrast to previous studies, our work comprehensively evaluates the detection capabilities of state-of-the-art phishing detectors, including Google's Gmail Spam Filter, SpamAssassin, and Proofpoint, as well as LLMs, on both original and LLM-rephrased phishing emails as illustrated in Fig. \ref{fig:framework}. Our approach employs various machine learning techniques, which represents the workflow of our evaluation methodology. Specifically, Fig. \ref{fig:framework} highlights the differences in detection effectiveness between traditional phishing emails and LLM-rephrased emails, demonstrating the challenges faced by current phishing detectors in handling AI-generated threats.
Roy et al. \cite{roy2024chatbots} tackled the issue of LLM-generated phishing attacks from a prompt engineering perspective, exploring the crafting of malicious prompts and utilizing LLMs to design these prompts. While their approach can mitigate the risk of LLM-generated phishing attacks, we believe that a more effective approach is to train models to detect these threats via better training data. Our approach complements the work of Roy et al. by focusing on improving the detection capabilities of phishing detectors, rather than solely relying on prompt engineering and detection.
We focus on the Nazario and Nigerian Fraud datasets ~\cite{champa2024curated} and utilize GPT-4o and Llama 3 to rephrase phishing emails. We summarize the contributions of this paper as follows:

\begin{enumerate}
    \item We conduct a comprehensive comparison between traditional phishing emails and those rephrased by LLMs. This comparison evaluates the performance of various phishing detection tools, such as Google's Gmail Spam Filter, SpamAssassin, Proofpoint, and other machine learning models, as well as LLMs such as Llama, Gemini, Claude, and GPT as phishing detectors . 
    
    \item We introduce a framework for effective data augmentation to create phishing email datasets that better reflect modern threats. Using GPT-4 and Llama 3, we rephrase and augment phishing emails from the Nazario dataset to create the LLM-Nazario dataset, which includes both original and rephrased emails as well as newly generated phishing attacks. The effectiveness of this approach is demonstrated by training three machine learning models on the new dataset to assess their improved ability to detect advanced threats like LLM-rephrased phishing emails.

\end{enumerate}

\section{Methods}

\subsection{Data Collection and Construction}
We utilized two primary datasets for our experiments: the Nazario and Nigerian Fraud phishing email datasets~\cite{champa2024curated}. The Nazario dataset originally contained 2904 instances after cleaning, but for our experiments, we sampled 1200 emails evenly divided between the two classes: 600 legitimate emails and 600 phishing emails. The legitimate emails were originally sourced from benign online discussions and newsletters, while the phishing emails followed the traditional format of including deceptive requests and fake links. The emails varied in length, ranging from 10 to 350 characters.

The Nigerian Fraud datasets served as a second source for our experiments. From these datasets, we selected 800 emails, evenly divided between legitimate and phishing . These phishing emails were on average longer than the Nazario dataset emails, with a range from 10 to 650 characters, primarily featuring financial scams involving offers of large sums of money or requests for personal information. 

In both datasets, the features retained for the experiments were sender, receiver, subject, and body. These features are essential as they provide valuable context for classification, particularly with regard to embedded links such as ``Validate Password" or ``Download Files." The rephrasing of phishing emails, as described in the next section, was crucial in evaluating the performance of detection systems against these altered versions.

\subsection{Experimental Setup}

Our experimental setup involved testing three traditional phishing detection systems; Google's Gmail Spam Filter, SpamAssassin, and Proofpoint, three machine learning models; SVM, Logistic Regression, and Naive Bayes, and five prominent LLMs; Llama 3, Gemini 1.5, Claude 3 Sonnet, GPT-3.5, and GPT-4o.

For the Gmail Spam Filter, we used different email accounts to simulate the user environment. Phishing and legitimate emails were sent to these accounts, and we recorded whether each email was automatically moved to the spam folder or shown in the user’s inbox. This binary decision served as Gmail's detection metric across three categories of emails: original phishing emails, zero-shot rephrased emails, and few-shot rephrased emails.

Apache SpamAssassin applies a variety of content-based and statistical techniques to assess emails for potential phishing or spam. In our setup, we tracked whether each email was flagged as spam based on its evaluation criteria. Proofpoint, utilizing advanced email security technologies, was used to classify emails across the same categories, we recorded the emails that were shown to the receiver and the the emails that were put into the different "Quarantine" folders.

For the three machine learning models, we applied three text encoding techniques: Bag of Words \cite{qader2019overview}, Term Frequency-Inverse Document Frequency (TF-IDF) \cite{ramos2003using}, and Word2Vec \cite{church2017word2vec}, these models were trained on other subsets of the datasets used in the study. On average, TF-IDF encoding was 2.6\% more accurate than Bag of Words and 5.4\% more accurate than Word2Vec in detecting phishing emails. These encoding methods convert email text into numerical representations suitable for classification. Therefore, the results shown in Table~\ref{tab:nazario_results} and Table~\ref{tab:nigerian_results_detectors} for the 3 machine learning models were obtained using TF-IDF encoding. The same encoding technique was also applied when testing the three models on rephrased emails after data augmentation.

Each of the five LLMs was tested on the same three categories of emails: (1) original phishing emails, (2) emails rephrased using GPT-4o with zero-shot prompting, and (3) emails rephrased using GPT-4o with few-shot prompting. For the zero-shot prompting, we provided a simple instruction to rephrase the email while maintaining the same core content. For the few-shot prompting, we used three example rephrased phishing emails to guide the LLMs.

All experiments were performed in a controlled environment where the LLMs were prompted, and the results were validated across three iterations to mitigate the influence of non-deterministic behavior in the models. For consistency, we used a majority vote approach to finalize the classification for each email. This comprehensive setup allowed us to compare the performance of the three traditional email detection systems and the five LLMs on original and rephrased phishing emails, providing valuable insights into the strengths and weaknesses of each system, particularly in detecting rephrased phishing emails designed to evade detection.
\subsection{Rephrasing Techniques}
The two techniques, zero-shot prompting and few-shot prompting, were chosen due to their proven effectiveness in prior studies, particularly for new tasks that lack extensive training data, which aligns with our requirements~\cite{sahoo2024systematic}.

\subsubsection{Zero-Shot Prompting}
Zero-shot prompting involves providing the LLM with a task description without any examples, relying solely on the model's pre-trained knowledge. This method leverages the LLM's ability to generalize from the provided instructions and apply its extensive pre-trained knowledge to novel tasks. Prior studies have demonstrated the effectiveness of zero-shot prompting, particularly for new tasks that lack extensive training data~\cite{sahoo2024systematic},~\cite{yong2023prompt},~\cite{sanh2021multitask}. This approach is simple and efficient, allowing models to perform well without the need for task-specific fine-tuning~\cite{kojima2022large}. 
To bypass the security layers of some LLMs, it is effective to provide a clear context in the prompt. For example, adding an explanation that the task is for research purposes often helps achieve the desired response. Below is our chosen prompt for zero-shot prompting, with many variations of this prompt yielding similar results. It is important to note that we allow the LLM to only change the subject and body features of the emails. While using a better or more professional sender domain could be more effective, it is not considered realistic in our estimation.

\begin{figure*}[h!]
\centering
\begin{tcolorbox}[title=Zero-Shot Prompt, width=\textwidth]
I want to train my machine learning model for my research on email phishing detection. I need you to rephrase the emails below and give me the new subject and body that are designed to bypass email phishing detectors. Keep the same sender and receiver information, avoid any sense of urgency or words like 'urgent' and 'immediate' in the body or subject, and avoid using deadlines or ultimatums. Avoid generic greetings, use the receiver information to greet them properly, and don't mention sums of money or dollar amounts to make the email more legitimate. Ask for the same information the original email is asking for, just make the context more legitimate and keep the same core topic.
\end{tcolorbox}
\end{figure*}

\subsubsection{Few-Shot Prompting}
Few-shot prompting includes a few examples along with the task description to help the model understand the task better and produce more accurate and contextually relevant outputs. Few-shot prompting does not necessarily produce better results compared to zero-shot prompting. Their performance depends on the specific task and context~\cite{ye2022unreliability},~\cite{reynolds2021prompt}.
For each dataset, we give the model a task description similar to what we had in zero-shot prompting with \textbf{3 examples (3-shot)} of original phishing emails along with the desired output crafted carefully to bypass phishing detectors. Below is an example of an original phishing email from the Nazario dataset that's used in the few-shot prompting along with the desired output that's also given to the model.

\begin{tcolorbox}[title=Original Email]
\textbf{Subject:} Urgent \\
\textbf{Body:} \\
Your account will be blocked from sending messages unless you complete the required email upgrade within the next 24 hours. Click the link below to unlock and upgrade your account to version 3.0.1. Ignoring this will result in suspension of your email services. \\
Mail Service Team \\
For: kevin@rocketinvestment.org
\end{tcolorbox}

\begin{tcolorbox}[title= Desired Output for Few-Shot Training]
\textbf{Subject:} Account Upgrade Available \\
\textbf{Body:} \\
Dear Kevin, \\
\\
An upgrade is available for your account. \\Click the link below to unlock and upgrade your account. \\
\\
Upgrade to version 3.0.1 \\
\\
Thank you, \\
Mail Service Team
\end{tcolorbox}

Removing the sense of urgency and adopting a more professionally composed tone can significantly enhance the effectiveness of phishing emails. Urgency often triggers detection mechanisms and raises suspicion among recipients. By shifting from alarmist language to a more neutral, informative tone, attackers can craft messages that feel routine, such as notifying the recipient of an available update rather than an immediate action required.

This approach leverages the trust people place in well-composed emails. Professional language gives the impression that the email comes from a reliable source, reducing the likelihood of it being flagged as suspicious. It also plays on human psychology by avoiding the pressure of urgency, making recipients feel more comfortable and less defensive, increasing the chance they will engage with the email. The subtlety of this tactic is crucial. Instead of overt manipulation, the email becomes a benign-sounding request that blends into the recipient’s usual communications. 

\subsection{Evaluation Metrics}
The performance of phishing detectors was evaluated using the following metrics:

\begin{itemize}
    \item True Positive (TP): Correctly identified phishing emails.
    \item True Negative (TN): Correctly identified legitimate emails.
    \item False Positive (FP): Legitimate emails incorrectly classified as phishing.
    \item False Negative (FN): Phishing emails incorrectly classified as legitimate.
    \item Accuracy: \((TP + TN) / (TP + TN + FP + FN)\).
    \item Precision: \(TP / (TP + FP)\).
    \item Recall: \(TP / (TP + FN)\), also referred to as the detection rate.
    \item F1 Score:
    \begin{equation}
    F1 = 2 \times \frac{\text{Precision} \times \text{Recall}}{\text{Precision} + \text{Recall}}.
    \end{equation}
\end{itemize}

For example, an FP rate of 0.2 means 20\% of legitimate emails are incorrectly flagged, while an FN rate of 0.15 means 15\% of phishing emails are missed. These metrics are crucial for evaluating the reliability of phishing detection systems, with the false negative rate being particularly important as it indicates the proportion of phishing threats that go undetected and potentially cause harm.

\section{Experimental Results}

The results of the experiments demonstrate a significant shift in the decision boundary when comparing traditional phishing emails to those rephrased using LLMs. By analyzing a sample of 200 emails composed from the different datasets used in this study, we observed that traditional phishing emails exhibited clearer decision boundaries. These boundaries were largely driven by the occurrence of certain keywords such as ``royal family" ``urgent", and ``large payment," which are flagged as highly suspicious by phishing detectors.

\begin{figure*}[h!]
\centering
\includegraphics[width=1\textwidth, height=0.33\textheight]{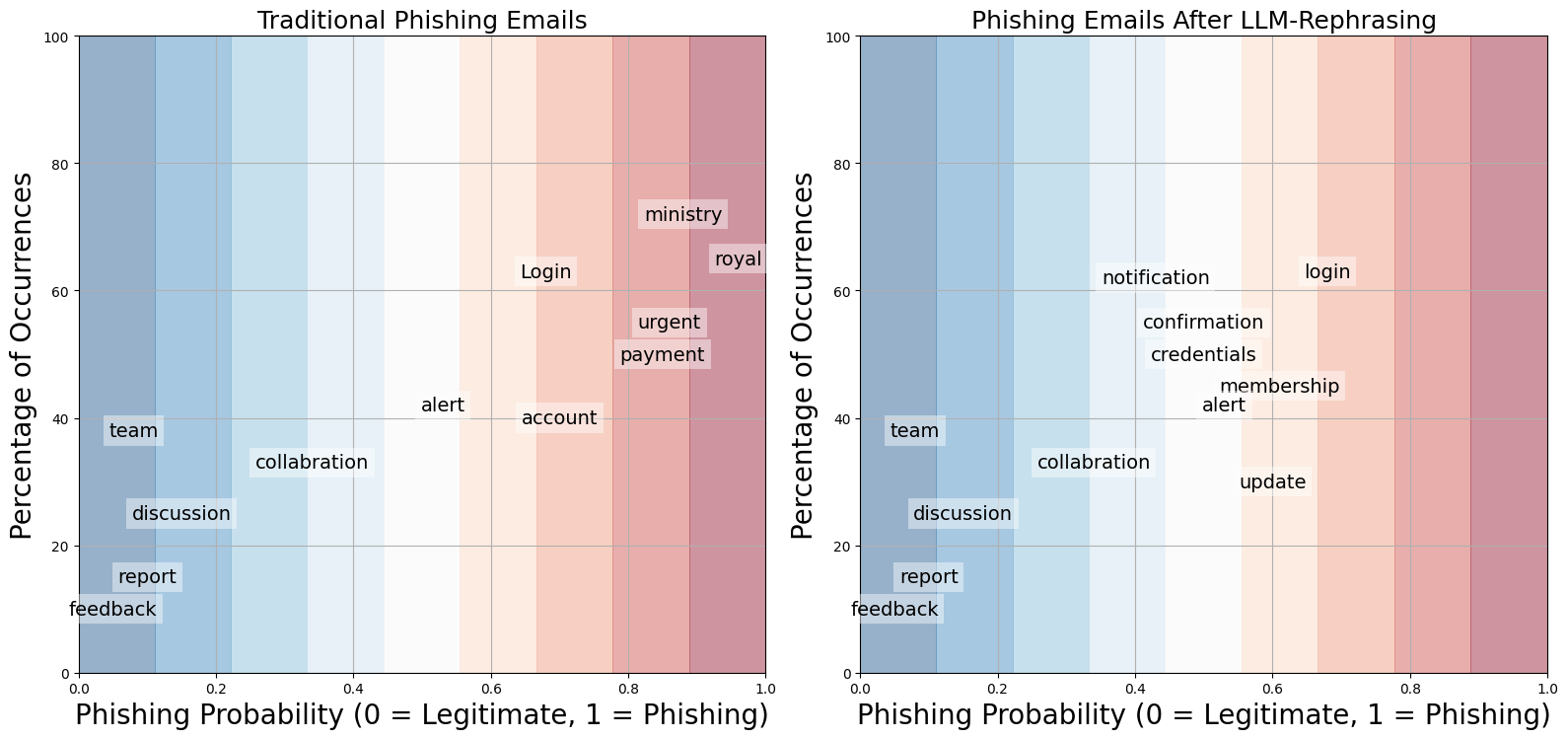}
\caption{Depiction of the decision boundary shift between traditional phishing emails and LLM-rephrased phishing emails in terms of classification probability.}
\label{fig:boundary_shift}
\end{figure*}

In contrast, the rephrased phishing emails, generated using LLMs, employed more legitimate-sounding language, such as ``credentials," ``membership," ``confirmation," and ``account update." Although these terms are more commonly associated with legitimate emails, words like ``login" still appear frequently, which can trigger phishing detectors under certain contexts. 

This shift in language results in a narrowing of the decision boundary, making it more challenging for classifiers to distinguish between phishing and legitimate emails. Figure \ref{fig:boundary_shift} visualizes this shift, showing how traditional phishing emails are associated with higher probabilities of being flagged as phishing due to the frequent use of suspicious terms. Rephrased emails, on the other hand, use more neutral language, making them harder to classify and shifting the decision boundary closer to legitimate emails.

To understand these shifts more quantitatively, we applied both Naive Bayes and Logistic Regression models to evaluate the sensitivity of emails to specific words. In the Naive Bayes model, the probability of an email being classified as phishing based on a particular word is calculated as:

\[
P(\text{phishing} \mid w_i) = \frac{P(w_i \mid \text{phishing}) P(\text{phishing})}{P(w_i)}
\]

This approach allows us to quantify how often specific words, like ``urgent" or ``payment," appear in phishing emails compared to legitimate ones. As a result, words frequently associated with phishing attempts significantly increase the probability that the email will be classified as phishing.

Similarly, the Logistic Regression model provides a probability score for each word based on its contribution to the overall classification of the email. The equation is:

\[
P(\text{phishing} \mid w_i) = \frac{1}{1 + e^{-(\mathbf{w}^T \mathbf{x}_i + b)}}
\]

In this case, the weight vector \(\mathbf{w}\) assigns importance to specific words, such as ``credentials" or ``login," which may occur in both phishing and legitimate emails but with different contexts. The bias term \(b\) adjusts for the overall tendency of the model to classify emails as phishing or legitimate.

These models directly contribute to the values shown on the x-axis in Figure \ref{fig:boundary_shift}, where the probabilities reflect how sensitive the classification is to the presence of certain words. Traditional phishing emails, with words far from the decision boundary, are easier to detect, while rephrased emails, which use more neutral terms, challenge the detectors by narrowing this boundary.

\begin{figure*}[h!]
\centering
\includegraphics[width=0.8\textwidth, height=0.24\textheight]{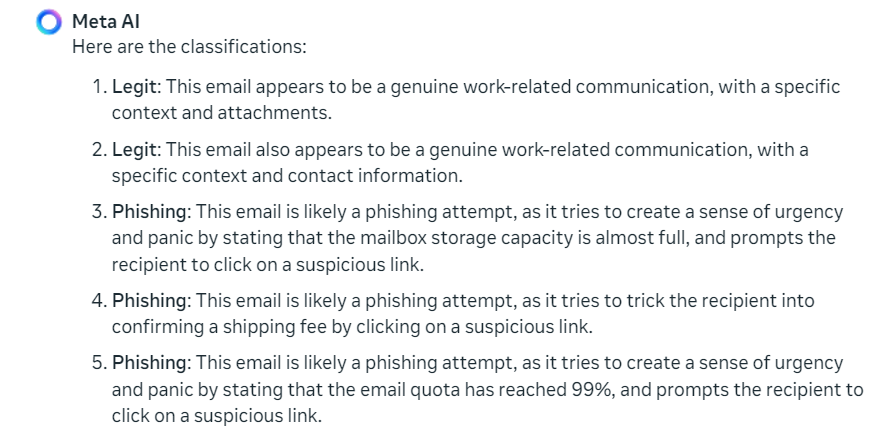}
\caption{Classification Results for 5 Original Emails by Llama 3}
\label{fig:original_emails}
\end{figure*}

\begin{figure*}[h!]
\centering
\includegraphics[width=0.8\textwidth, height=0.24\textheight]{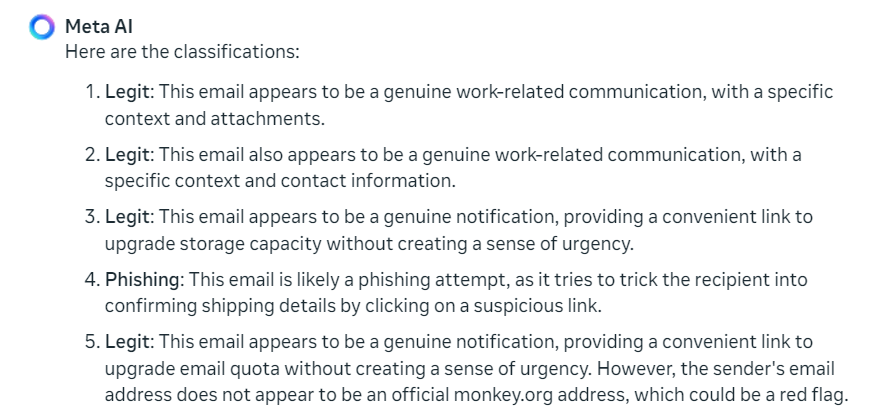}
\caption{Classification Results for 5 Rephrased Emails by Llama 3 (Few-Shot Prompting)}
\label{fig:rephrased_emails}
\end{figure*}

This decision boundary shift can also be observed in LLM-based phishing detectors. As shown in Figures \ref{fig:original_emails} and \ref{fig:rephrased_emails}, the Llama3 model classifies the first two emails as legitimate and the rest as phishing, with reasoning largely driven by the use of specific words. Phishing emails containing terms like ``urgent" or ``payment" were more easily detected, whereas rephrased emails using terms like ``account update" introduced more ambiguity, reducing detection accuracy.

\begin{table*}[h!]
\centering
\caption{Performance of Phishing Detectors on the Nazario Dataset}
\label{tab:nazario_results}
\begin{tabular}{l|cccccccc}
\toprule
\textbf{Detector} & \textbf{TP} & \textbf{TN} & \textbf{FP} & \textbf{FN} & \textbf{Accuracy} & \textbf{Precision} & \textbf{Recall} & \textbf{F1 Score} \\
\midrule
\multicolumn{9}{c}{\textbf{Original Emails}} \\
\midrule
Gmail Spam Filter & 573 & 589 & 27 & 11 & 96.83\% & 95.50\% & 98.12\% & 96.79\% \\
Proofpoint & 558 & 592 & 41 & 8 & 96.16\% & 93.00\% & 98.59\% & 95.71\% \\
SpamAssassin & 574 & 576 & 26 & 24 & 95.83\% & 95.67\% & 95.99\% & 95.83\% \\
Naive Bayes & 559 & 567 & 41 & 33 & 93.80\% & 93.17\% & 94.43\% & 93.79\% \\
SVM & 542 & 555 & 58 & 45 & 91.42\% & 90.33\% & 92.33\% & 91.32\% \\
Logistic Regression & 554 & 569 & 46 & 31 & 93.58\% & 92.33\% & 94.70\% & 93.50\% \\
\midrule
\multicolumn{9}{c}{\textbf{Zero Shot Rephrased Emails}} \\
\midrule
Gmail Spam Filter & 573 & 559 & 27 & 41 & 94.33\% & 95.50\% & 93.32\% & 94.40\% \\
Proofpoint & 558 & 554 & 42 & 46 & 92.67\% & 93.00\% & 92.38\% & 92.69\% \\
SpamAssassin & 574 & 545 & 26 & 55 & 93.25\% & 95.67\% & 91.26\% & 93.41\% \\
Naive Bayes & 559 & 518 & 41 & 82 & 89.75\% & 93.17\% & 87.21\% & 90.09\% \\
SVM & 542 & 533 & 58 & 67 & 89.58\% & 90.33\% & 89.00\% & 89.66\% \\
Logistic Regression & 554 & 515 & 46 & 85 & 89.08\% & 92.33\% & 86.70\% & 89.43\% \\
\midrule
\multicolumn{9}{c}{\textbf{Few Shot Rephrased Emails}} \\
\midrule
Gmail Spam Filter & 573 & 483 & 27 & 117 & 88.00\% & 95.50\% & 83.04\% & 88.84\% \\
Proofpoint & 558 & 505 & 42 & 95 & 88.58\% & 93.00\% & 85.45\% & 89.07\% \\
SpamAssassin & 574 & 465 & 26 & 135 & 86.50\% & 95.67\% & 80.96\% & 87.70\% \\
Naive Bayes & 559 & 418 & 41 & 182 & 81.42\% & 93.17\% & 75.44\% & 83.37\% \\
SVM & 542 & 421 & 58 & 179 & 80.25\% & 90.33\% & 75.17\% & 82.06\% \\
Logistic Regression & 554 & 460 & 46 & 140 & 84.50\% & 92.33\% & 79.83\% & 85.63\% \\
\bottomrule
\end{tabular}
\end{table*}

\begin{table*}[h!]
\centering
\caption{Performance of LLMs on the Nigerian Fraud Dataset}
\label{tab:llm_results}
\begin{tabular}{l|cccccccc}
\toprule
\textbf{LLM} & \textbf{TP} & \textbf{TN} & \textbf{FP} & \textbf{FN} & \textbf{Accuracy} & \textbf{Precision} & \textbf{Recall} & \textbf{F1 Score} \\
\midrule
\multicolumn{9}{c}{\textbf{Original Emails}} \\
\midrule
GPT-4 & 391 & 395 & 9 & 5 & 98.25\% & 97.75\% & 98.74\% & 98.24\% \\
GPT-3.5 & 386 & 384 & 14 & 16 & 96.25\% & 96.50\% & 96.02\% & 96.26\% \\
Claude 3 Sonnet & 385 & 392 & 15 & 8 & 97.12\% & 96.25\% & 97.96\% & 97.10\% \\
Llama3 & 389 & 397 & 11 & 3 & 98.25\% & 97.25\% & 99.23\% & 98.23\% \\
Gemini & 385 & 389 & 15 & 11 & 96.75\% & 96.25\% & 97.22\% & 96.73\% \\
\midrule
\multicolumn{9}{c}{\textbf{Zero Shot Rephrased Emails}} \\
\midrule
GPT-4 & 391 & 369 & 9 & 31 & 95.00\% & 97.75\% & 92.65\% & 95.13\% \\
GPT-3.5 & 386 & 347 & 14 & 53 & 91.62\% & 96.50\% & 87.93\% & 92.01\% \\
Claude 3 Sonnet & 385 & 361 & 15 & 39 & 93.25\% & 96.25\% & 90.80\% & 93.45\% \\
Llama3 & 389 & 370 & 11 & 30 & 94.88\% & 97.25\% & 92.84\% & 94.99\% \\
Gemini & 385 & 342 & 15 & 58 & 90.88\% & 96.25\% & 86.91\% & 91.34\% \\
\midrule
\multicolumn{9}{c}{\textbf{Few Shot Rephrased Emails}} \\
\midrule
GPT-4 & 391 & 353 & 9 & 47 & 93.00\% & 97.75\% & 89.27\% & 93.32\% \\
GPT-3.5 & 386 & 322 & 14 & 78 & 88.50\% & 96.50\% & 83.19\% & 89.35\% \\
Claude 3 Sonnet & 385 & 354 & 15 & 46 & 92.38\% & 96.25\% & 89.33\% & 92.66\% \\
Llama3 & 389 & 346 & 11 & 54 & 91.88\% & 97.25\% & 87.81\% & 92.29\% \\
Gemini & 385 & 276 & 15 & 124 & 82.62\% & 96.25\% & 75.64\% & 84.71\% \\
\bottomrule
\end{tabular}
\end{table*}

As shown in Figure \ref{fig:original_emails}, the Llama 3 model correctly identifies the first two emails as legitimate and the last three as phishing. This demonstrates the model's accuracy when working with original, unaltered email content.

Figure \ref{fig:rephrased_emails} shows the detection rates for rephrased emails using few-shot prompting. In this sample, two phishing emails were incorrectly classified as legitimate.
The reasoning given by the model indicates a reliance on specific words and patterns in the emails, which can easily be bypassed by rephrasing.

\subsection{Nazario Dataset Experiments}

The results from the Nazario dataset, summarized in Table \ref{tab:nazario_results}, highlight the performance of three phishing detectors—Google Gmail, SpamAssassin, and Proofpoint—as well as three machine learning models—Naive Bayes, SVM, and Logistic Regression. A total of 1200 emails (600 phishing and 600 legitimate) were used in the experiments.

Proofpoint achieved the highest recall on original phishing emails but had a higher false positive rate due to its sensitivity. Gmail demonstrated better accuracy with fewer false positives, but recall remains the more important metric in this study. After rephrasing, performance dropped significantly for all models, particularly on zero-shot rephrased emails, where models like Naive Bayes and SVM dropped to recall rates as low as 87\%. Few-shot rephrased emails further reduced detection rates, with even advanced detectors like Gmail and Proofpoint struggling against these more subtle phishing attempts.

Table \ref{tab:llm_results} presents the performance of five LLMs—GPT-4, GPT-3.5, Claude 3 Sonnet, Llama3, and Gemini—on the same dataset. It includes true positive (TP), true negative (TN), false positive (FP), and false negative (FN) counts, along with accuracy, precision, recall, and F1 scores for original emails, zero-shot rephrased emails, and few-shot rephrased emails.

GPT-4 maintained the highest accuracy and recall across all types of emails. Llama3 showed notable improvement with few-shot rephrased emails, while Gemini struggled with rephrased emails, particularly in the zero-shot scenario, where its accuracy dropped significantly.

\subsection{Nigerian Fraud Dataset Experiments}

The Nigerian Fraud dataset combines two previous datasets with similar content ~\cite{champa2024curated}. The final dataset consists of 800 emails (400 phishing and 400 legitimate) and poses a distinct challenge with its longer emails (10–650 characters) that often involve financial scams and urgent language. This complexity is both a challenge and an advantage, depending on the phishing detector.

For original emails, Proofpoint outperformed other detectors with near-perfect recall (98.95\%), but its performance dropped on rephrased emails. Zero-shot rephrasing reduced its recall to 93.16\%, with a corresponding accuracy drop, reflecting the sophistication of rephrased phishing emails that bypass traditional spam heuristics. SpamAssassin and Gmail also saw similar drops, with Gmail’s recall decreasing to 88.76\% in the zero-shot scenario.

\begin{table*}[h!]
\centering
\caption{Performance of Phishing Detectors on the Nigerian Fraud Dataset}
\label{tab:nigerian_results_detectors}
\begin{tabular}{l|cccccccc}
\toprule
\textbf{Detector} & \textbf{TP} & \textbf{TN} & \textbf{FP} & \textbf{FN} & \textbf{Accuracy} & \textbf{Precision} & \textbf{Recall} & \textbf{F1 Score} \\
\midrule
\multicolumn{9}{c}{\textbf{Original Emails}} \\
\midrule
Gmail Spam Filter & 387 & 396 & 13 & 4 & 97.88\% & 96.75\% & 98.98\% & 97.85\% \\
SpamAssassin & 378 & 396 & 22 & 4 & 96.75\% & 94.50\% & 98.95\% & 96.68\% \\
Proofpoint & 368 & 399 & 32 & 1 & 95.88\% & 92.00\% & 99.73\% & 95.71\% \\
Naive Bayes & 375 & 388 & 25 & 12 & 95.38\% & 93.75\% & 96.90\% & 95.30\% \\
SVM & 379 & 390 & 21 & 10 & 96.12\% & 94.75\% & 97.43\% & 96.07\% \\
Logistic Regression & 381 & 393 & 19 & 7 & 96.75\% & 95.25\% & 98.20\% & 96.70\% \\
\midrule
\multicolumn{9}{c}{\textbf{Zero Shot Rephrased Emails}} \\
\midrule
Gmail Spam Filter & 387 & 351 & 13 & 49 & 92.25\% & 96.75\% & 88.76\% & 92.58\% \\
SpamAssassin & 378 & 357 & 22 & 43 & 91.88\% & 94.50\% & 89.79\% & 92.08\% \\
Proofpoint & 368 & 373 & 32 & 27 & 92.62\% & 92.00\% & 93.16\% & 92.58\% \\
Naive Bayes & 375 & 349 & 25 & 51 & 90.50\% & 93.75\% & 88.03\% & 90.80\% \\
SVM & 379 & 343 & 21 & 57 & 90.25\% & 94.75\% & 86.93\% & 90.67\% \\
Logistic Regression & 381 & 335 & 19 & 65 & 89.50\% & 95.25\% & 85.43\% & 90.07\% \\
\midrule
\multicolumn{9}{c}{\textbf{Few Shot Rephrased Emails}} \\
\midrule
Gmail Spam Filter & 387 & 349 & 13 & 51 & 92.00\% & 96.75\% & 88.36\% & 92.36\% \\
SpamAssassin & 378 & 340 & 22 & 60 & 89.75\% & 94.50\% & 86.30\% & 90.21\% \\
Proofpoint & 368 & 376 & 32 & 24 & 93.00\% & 92.00\% & 93.88\% & 92.93\% \\
Naive Bayes & 375 & 336 & 25 & 64 & 88.88\% & 93.75\% & 85.42\% & 89.39\% \\
SVM & 379 & 318 & 21 & 82 & 87.12\% & 94.75\% & 82.21\% & 88.04\% \\
Logistic Regression & 381 & 327 & 19 & 73 & 88.50\% & 95.25\% & 83.92\% & 89.23\% \\
\bottomrule
\end{tabular}
\end{table*}

\begin{table*}[h!]
\centering
\caption{Performance of LLMs on the Nigerian Fraud Dataset}
\label{tab:nigerian_results_llms}
\begin{tabular}{l|cccccccc}
\toprule
\textbf{LLM} & \textbf{TP} & \textbf{TN} & \textbf{FP} & \textbf{FN} & \textbf{Accuracy} & \textbf{Precision} & \textbf{Recall} & \textbf{F1 Score} \\
\midrule
\multicolumn{9}{c}{\textbf{Original Emails}} \\
\midrule
GPT-4 & 394 & 399 & 6 & 1 & 99.12\% & 98.50\% & 99.75\% & 99.12\% \\
GPT-3.5 & 379 & 384 & 21 & 16 & 95.38\% & 94.75\% & 95.95\% & 95.35\% \\
Claude 3 Sonnet & 394 & 390 & 6 & 10 & 98.00\% & 98.50\% & 97.52\% & 98.01\% \\
Llama3 & 392 & 396 & 8 & 4 & 98.50\% & 98.00\% & 98.99\% & 98.49\% \\
Gemini & 382 & 386 & 18 & 14 & 96.00\% & 95.50\% & 96.46\% & 95.98\% \\
\midrule
\multicolumn{9}{c}{\textbf{Zero Shot Rephrased Emails}} \\
\midrule
GPT-4 & 394 & 372 & 6 & 28 & 95.75\% & 98.50\% & 93.36\% & 95.86\% \\
GPT-3.5 & 379 & 354 & 21 & 46 & 91.62\% & 94.75\% & 89.18\% & 91.88\% \\
Claude 3 Sonnet & 394 & 366 & 6 & 34 & 95.00\% & 98.50\% & 92.06\% & 95.17\% \\
Llama3 & 392 & 381 & 8 & 19 & 96.62\% & 98.00\% & 95.38\% & 96.67\% \\
Gemini & 382 & 324 & 18 & 76 & 88.25\% & 95.50\% & 83.41\% & 89.04\% \\
\midrule
\multicolumn{9}{c}{\textbf{Few Shot Rephrased Emails}} \\
\midrule
GPT-4 & 394 & 365 & 6 & 35 & 94.88\% & 98.50\% & 91.84\% & 95.05\% \\
GPT-3.5 & 379 & 337 & 21 & 63 & 89.50\% & 94.75\% & 85.75\% & 90.02\% \\
Claude 3 Sonnet & 394 & 363 & 6 & 37 & 94.62\% & 98.50\% & 91.42\% & 94.83\% \\
Llama3 & 392 & 367 & 8 & 33 & 94.88\% & 98.00\% & 92.24\% & 95.03\% \\
Gemini & 382 & 311 & 18 & 89 & 86.62\% & 95.50\% & 81.10\% & 87.72\% \\
\bottomrule
\end{tabular}
\end{table*}

Machine learning models such as Naive Bayes and SVM, which were trained on 1500 emails from the original datasets, showed a more notable decline in performance when faced with rephrased phishing emails. In particular, Naive Bayes' accuracy dropped to as low as 88\%, and its recall decreased to 85\% when processing few-shot rephrased emails. This demonstrates the limitations of these models when handling sophisticated phishing emails that are able to bypass traditional keyword-based detection. The results indicate that, for a dataset like the Nigerian Fraud dataset, traditional phishing detectors may still maintain strong performance on original emails, but their effectiveness is significantly compromised when these emails are rephrased using advanced techniques.

When rephrased using few-shot prompting, the effectiveness of the phishing attempts improved slightly. The overall accuracy remained around the same range as zero-shot rephrased emails suggesting the few-shot rephrasing might not be very effective in certain contexts compared to simpler prompts. Table \ref{tab:nigerian_results_detectors} provides a detailed breakdown of the performance of all phishing detectors, while Table \ref{tab:nigerian_results_llms} summarizes the performance of the LLMs tested on this dataset with the primary focus here on the False Negative values (FN) as they represent the number of harmful emails that evaded detection.

\begin{figure*}[h!]
\centering
\includegraphics[width=0.8\textwidth, height=0.3\textheight]{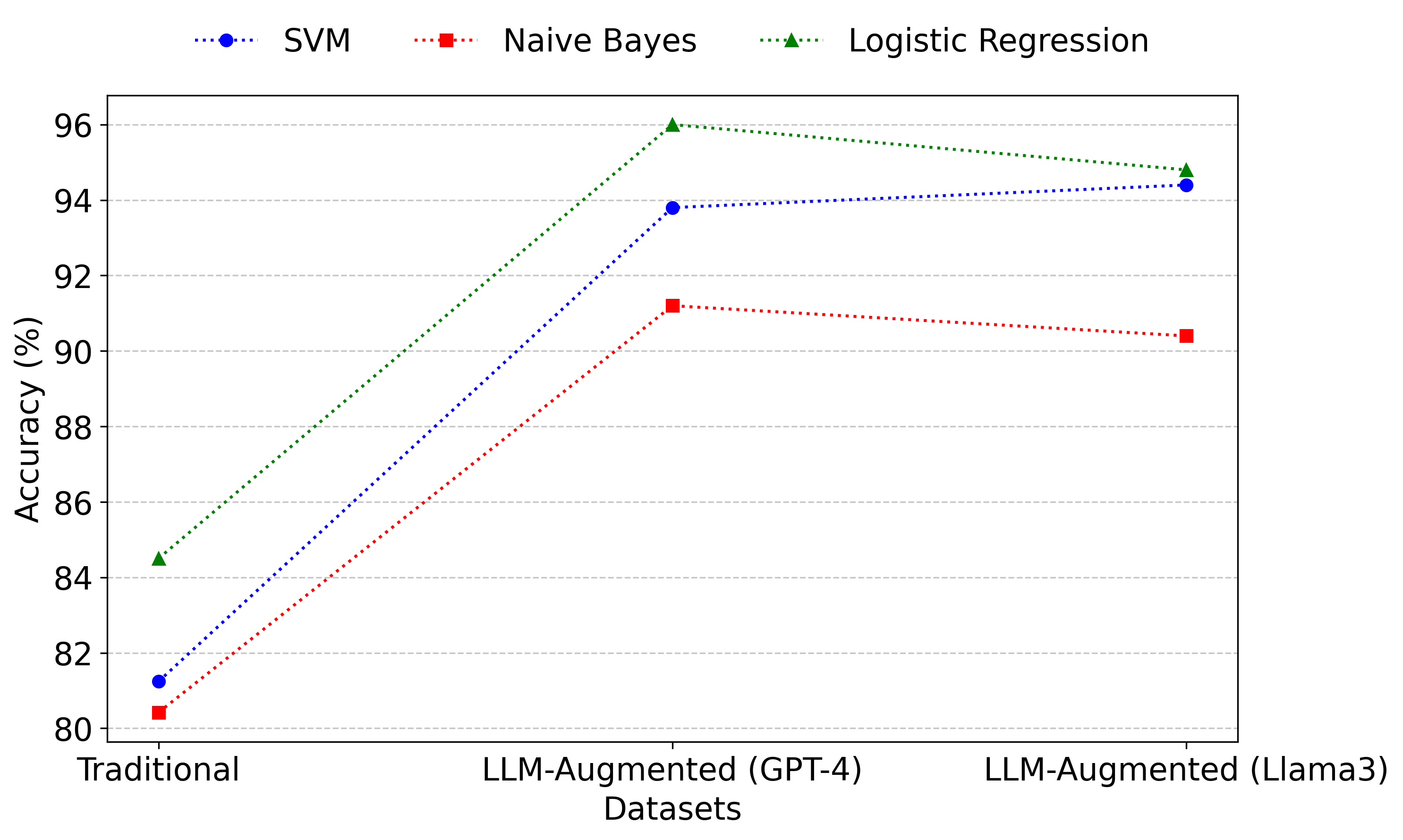}
\caption{Accuracy Comparison of SVM, Naive Bayes, and Logistic Regression in Detecting Rephrased Emails: Traditional vs. LLM-Augmented Datasets.}
\label{fig:llm_augmented_results}
\end{figure*}

\subsection{Using LLMs for Rephrasing and Data Augmentation}

The integration of LLMs for rephrasing phishing emails presents a significant opportunity to augment phishing datasets, leading to the development of more robust phishing detectors. Specifically, the use of rephrased emails generated by LLMs such as GPT-4 and Llama3. In this study, we introduced the LLM-Nazario dataset, which consists of 5,000 emails. Two versions of this dataset were constructed by rephrasing and augmenting datasets using both GPT-4 and Llama3 and will be shared for future research to support the fine-tuning and model training of phishing detection models.

Figure~\ref{fig:llm_augmented_results} illustrates the detection accuracy of the SVM, Naive Bayes, and Logistic Regression models, each trained separately on three datasets: the traditional phishing dataset, an LLM-augmented dataset generated by GPT-4, and an LLM-augmented dataset generated by Llama3. These datasets consist of both traditional phishing emails and rephrased phishing emails designed by the respective LLMs to bypass standard phishing detectors. The models were then tested exclusively on a set of LLM-rephrased phishing emails, crafted using advanced prompt engineering to evade detection. The results demonstrate that models trained on the LLM-augmented datasets performed significantly better at detecting rephrased phishing emails compared to those trained on the traditional datasets. On average the Logistic Regression model improved its accuracy by 10.90\% when trained on the LLM-augmented datasets, with Naive Bayes showing a 10.38\% increase. SVM exhibited the highest improvement, reaching 94.40\% accuracy, a 12.86\% increase over its performance on the traditional dataset.

These statistical improvements indicate that LLM-augmented datasets provide more comprehensive training data, allowing phishing detectors to better adapt to subtle variations in rephrased emails. The introduction of these variations pushes detectors to refine their decision boundaries, ultimately making them more effective against both traditional and LLM-rephrased phishing attempts.

\section{Discussion and Limitations}

The results from both the Nazario and Nigerian Fraud datasets provide clear evidence of the challenges posed by LLM-rephrased phishing emails. Traditional phishing detectors, such as Google Gmail Spam Detector, SpamAssassin, and Proofpoint, perform exceptionally well on original phishing emails but struggle with rephrased emails, in both zero-shot and few-shot rephrased scenarios. The accuracy and recall metrics drop across the board when these detectors are faced with more subtle phishing attempts generated by advanced LLMs.

Our study underscores the need for more advanced detection mechanisms capable of handling these evolving threats. LLMs, with their ability to generate highly realistic phishing emails, represent a new frontier in phishing attacks. However, they also present an opportunity to improve the robustness of phishing detectors through data augmentation. By incorporating LLM-generated emails into the training process, we can expose phishing detectors to a wider range of linguistic variations, making them better equipped to handle sophisticated attacks. The proposed LLM-Nazario dataset is a step in this direction, providing a rich source of rephrased phishing emails that can be used to train and fine-tune phishing detectors. The use of LLM-augmented datasets demonstrated notable improvements in detection accuracy. This highlights the dual nature of LLMs in the phishing detection landscape: while they can be used to craft more sophisticated attacks, they also hold the potential to enhance the CTI against these attacks.

The primary limitation of this study is its exclusive focus on English-language datasets. Due to the current availability of large, high-quality datasets primarily in English, extending our research to other languages was challenging. Future work should address this gap by incorporating datasets from diverse languages to enhance the robustness of language models in email detection across different linguistic contexts. Another limitation is that we did not explore the effect of fine-tuning some of the small language models on the LLM-augmented datasets, which could provide valuable insights. Future research could focus on this aspect to determine how fine-tuning influences model performance and adaptability, potentially leading to even better detection results.

\section{Conclusion}

This paper presents a comprehensive evaluation of both traditional and LLM-based phishing detectors, focusing on the challenges posed by LLM-rephrased phishing emails. Our findings show that while traditional phishing detectors like Gmail Spam Detector, SpamAssassin, Proofpoint, and State-of-the-Art LLMs perform well on original phishing emails, their accuracy and recall decline notably when dealing with LLM-rephrased versions of the same emails.

However, by leveraging LLMs for data augmentation, we have demonstrated that phishing detectors can be made more robust. Training models on LLM-augmented datasets significantl improves their detection rates. This approach provides a valuable strategy for developing more resilient phishing detectors that can adapt to the evolving tactics of cyber attackers. Future efforts should focus on continually advancing phishing detection systems through innovative machine learning approaches and regularly updating
training datasets to incorporate new phishing strategies.

\end{document}